\documentclass{aip-cp}

\usepackage[numbers]{natbib}
\usepackage{rotating}
\usepackage{graphicx}

\def\be{\begin{equation}}
\def\ee{\end{equation}}
\def\bea{\begin{eqnarray}}
\def\eea{\end{eqnarray}}

\def\mev{\, {\rm MeV}}

\newcommand{\gsim}{\lower.7ex\hbox{$\;\stackrel{\textstyle>}{\sim}\;$}}
\newcommand{\lsim}{\lower.7ex\hbox{$\;\stackrel{\textstyle<}{\sim}\;$}}

\newcommand{\cm}{\rm cm}
\newcommand{\s}{\rm s}
\newcommand{\sr}{\rm sr}

\graphicspath{ {../Figures/} }

\begin{document}

\title{Minding the MeV Gap: the Indirect Detection of Low Mass Dark Matter }

\author[aff1]{Kimberly K.~Boddy}
\author[aff1]{Jason Kumar\corref{cor1}}

\affil[aff1]{Department of Physics and Astronomy, University of Hawai'i,
  Honolulu, HI 96822, USA}
\corresp[cor1]{Corresponding author: jkumar@hawaii.edu}

\maketitle

\begin{abstract}
We consider the prospects for the indirect detection of low mass dark matter which couples dominantly
to quarks.  If the center of mass energy is below about 280 MeV, the kinematically allowed final states
will be dominated by photons and neutral pions, producing striking signatures at gamma ray telescopes.
In fact, an array of new instruments have been proposed, which would greatly improve sensitivity to photons
in this energy range.  We find that planned instruments can improve on current sensitivity to dark matter
models of this type by up to a few orders of magnitude.
\end{abstract}

\section{INTRODUCTION}

There are three reasons to focus attention on indirect detection signatures in the photon spectrum arising from
the annihilation or decay of very low-mass dark matter ($m_X \sim {\cal O}(100\mev)$)~\cite{Boddy:2015efa}:
\begin{itemize}
  \item{If dark matter couples dominantly to quarks and is sufficiently light, then the annihilation/decay
  products are dominated by $\gamma$s and $\pi^0$s, producing striking signatures in the photon spectrum; }
  \item{A variety of new astronomical instruments which can vastly improve sensitivity to ${\cal O}(1-100)\mev$
  photons  are under consideration;}
  \item{Direct detection experiments tend to lose sensitivity at such low masses, while indirect
  detection strategies may well be the most effective.}
\end{itemize}

For a dark matter initial state with center-of-mass energy $\sqrt{s} < 2m_{\pi^\pm}$, the only kinematically
accessible non-leptonic two-body final states are $\gamma \gamma$, $\gamma \pi^0$ and $\pi^0 \pi^0$.  Since the $\pi^0$ will
decay almost exclusively to $\gamma \gamma$, low-mass dark matter is essentially ``guaranteed" to produce interesting
indirect detection signatures in the photon spectrum, provided it couples to quarks.
Moreover, since sensitivity to photons in this energy range is relatively poor, there are a variety of proposals for
new instruments which would dramatically improve sensitivity in this ``MeV gap,"~\cite{Greiner:2011ih} including
the Advanced Compton Telescope (ACT)~\cite{Boggs:2006mh}, the Advanced Pair Telescope (APT)~\cite{Buckley:2008fk}, the
Gamma-Ray Imaging, Polarimetry and Spectroscopy (GRIPS) detector~\cite{Greiner:2011ih}, the Advanced Energetic Pair Telescope
(AdEPT)~\cite{Hunter:2013wla}, the Pair-Production Gamma-Ray Unit (PANGU)~\cite{Wu:2014tya}, the Compton Spectrometer and
Imager (COSI)~\cite{Kierans:2014lr}, and ASTROGAM~\cite{astrogam}.
In this proceedings contribution, we describe the bounds on dark matter annihilation or decay to these final states
arising from current data (from COMPTEL, EGRET and Fermi), and the future sensitivity of other instruments under consideration.

We begin by describing a class of low-mass dark matter models which produce distinctive photon spectra.
We then describe the general features of indirect detection strategies  for probing these models,
utilizing either the diffuse gamma ray flux or a
search of dwarf galaxies.  We describe the current bounds arising from existing gamma ray data sets in this energy range (as well as
from the Cosmic Microwave Background (CMB)), and the sensitivity of future contemplated instruments.  We end with a discussion of our
results.

\section{MODELS AND SPECTRA}

We consider the case of low-mass dark matter whose dominant Standard Model coupling is to quarks.  Indirect detection is possible
through either dark matter annihilation or decay; in either case, we refer to the center-of-mass energy of the relevant process
as $\sqrt{s}$, and assume $\sqrt{s} < 2m_{\pi^\pm}$.  The only non-leptonic two-body final states that are
kinematically allowed are $\gamma \gamma$, $\gamma \pi^0$, and $\pi^0 \pi^0$.  Additional leptonic or three-body final states are also
available, but the relevant annihilation/decay rates are suppressed by factors of $\alpha$ or $sG_F$, and we assume they are negligible.

For each two-body final state, the energies of the prompt photons and neutral pions are determined by $\sqrt{s}$.  The
neutral pion decays through the process $\pi^0 \rightarrow \gamma \gamma$ with $\sim 99\% $ efficiency,
resulting in a ``box-like" spectrum, characteristic
of monoenergetic photons from an isotropically boosted source.  As a result, all
of the final state products produce distinctive photon signatures.

If we assume that weak interactions are negligible and that the fundamental dark matter-quark interaction is $C$-invariant, then we may
classify final states by their $C$ eigenvalues.  The photon injection spectrum associated with each final state is determined by kinematics, and are
as follows:
\begin{itemize}
\item $\pi^0 \pi^0$ ($C$-even):
  \bea
  \frac{dN_\gamma}{dE} &=& \frac{4}{\Delta E} \left[\Theta(E-E_-)-\Theta(E - E_+)\right] ,
  \nonumber\\
    E_\pm &=& \frac{\sqrt{s}}{4} \pm {\Delta E \over 2}  \ , \qquad
    \Delta E = {\sqrt{s} \over 2}\sqrt{1- {4m_{\pi^0}^2 \over s} } \ .
  \eea
\item $\gamma \pi^0$ ($C$-odd):
  \bea
  \frac{dN_\gamma}{dE} &=& \delta (E-E_0) + \frac{2}{\Delta E} \left[\Theta(E-E_-)-\Theta(E - E_+)\right] ,
  \nonumber\\
    E_0 &=& \frac{\sqrt{s}}{2} \left(1-\frac{m_{\pi^0}^2}{s}\right)
    \nonumber\\
  E_\pm &=& \frac{\sqrt{s}}{4} \left(1+\frac{m_{\pi^0}^2}{s}\right)  \pm {\Delta E \over 2}  \ , \qquad
  \Delta E = \frac{\sqrt{s}}{2} \left(1-\frac{m_{\pi^0}^2}{s} \right) \ .
\eea
\item $\gamma\gamma$ ($C$-even):
  \bea
  \frac{dN_\gamma}{dE} &=& 2\delta (E-E_0)
  \nonumber\\
    E_0 &=& \frac{\sqrt{s}}{2} \ .
  \eea
\end{itemize}
Depending on the details of the model, as well as the $C$ eigenvalue and energy of the initial state, it is possible for any final
state to dominate.  We will therefore analyze the detection prospects for each final state channel assuming the branching fraction
to that channel is 1.

The above photon injection spectra must be convolved with an energy resolution function to determine the spectrum observed at
any particular detector.  For simplicity, we will assume in our analysis that the energy resolution function for each detector
can be well-approximated by a Gaussian with an energy-independent fractional width.

For channels in which a $\pi^0$ is produced just above threshold, the $\pi^0$ will decay nearly at rest, producing a very narrow
box-like feature ($\Delta E$ small) which may mimic a line.  This feature will provide an excellent signature for a detector
whose energy range is large enough to contain the peak.  In the $\gamma \pi^0$ channel, this box-like feature will dominate the sensitivity
of an experiment only if it is sufficiently narrow; for large enough $\sqrt{s}$, the monoenergetic photon will be a larger driver of sensitivity.

\section{SEARCHES OF THE DIFFUSE FLUX AND OF DWARF GALAXIES}

We will focus on signals of dark matter annihilation/decay in the diffuse gamma ray flux, and in
gamma ray emission from dwarf spheroidal galaxies.  This analysis is described in detail in~\cite{Boddy:2015efa}.
The diffuse flux of gamma rays has been measured in the energy range of
interest by COMPTEL and EGRET~\cite{Strong:2004de}.  The data are relatively smooth and can be fit to a power law~\cite{Boddy:2015efa}:
\begin{equation}
  d^2 \Phi / dE\, d\Omega = (2.74 \times 10^{-3}  ~\cm^{-2}\s^{-1}\sr^{-1}\mev^{-1}) ( E /\mev )^{-2.0} \ .
  \label{eq:bkg-fit}
\end{equation}
A indirect detection search in the diffuse gamma ray flux is essentially a search for a sharp feature
in the observed photon spectrum.  On the other hand, the observed diffuse photon spectrum constitutes the
foreground to a search for dark matter in dwarf galaxies.

To determine the sensitivity of current and future instruments to searches for diffuse gamma ray emission,
we consider both a ``conservative" and ``optimistic" analysis.  In the conservative analysis, we exclude
models for which the expected number of photons from dark matter annihilation/decay exceeds the observed
number of counts by $2\sigma$ in any one energy bin.\footnote{In determining bounds from current data,
we utilize the actual flux measurements and energy binning of the relevant instrument.  In estimating the sensitivity
of future experiments, we assume that the observed flux is consistent with Equation~\ref{eq:bkg-fit}, and adopt an
optimal energy binning for the relevant experiment, given an estimate for its energy resolution.}
This analysis would be appropriate if there is little
confidence in one's understanding of the underlying astrophysical backgrounds.  But in a more optimistic case,
one might be confident that the underlying astrophysical background exhibited no sharp features.  In that case,
one could instead estimate the systematic uncertainty in a smooth fit to the diffuse gamma ray flux (following
Reference~\cite{Strong:2004de},
we estimate the systematic uncertainty to be 15\%), and exclude models for which the expected number of photons from the dark matter
signal exceeded twice this uncertainty in any one energy bin.

It is important to note that, for either a conservative or optimistic analysis of the diffuse
gamma ray flux, statistical uncertainties are subleading.  As a result, sensitivity is not significantly
affected by the exposure or angular resolution of the instrument.  Instead, the sensitivity to a line signal is
proportional to $\epsilon^{-1}$, which determines the instrument's ability to distinguish a sharp feature in the data.

For an analysis of dwarf galaxies, however, we assume that it is possible to estimate the foreground photon spectrum
arising from astrophysical sources (and from diffuse gamma ray emission due dark matter annihilation/decay along the line
of sight) by looking slightly
off-axis.  One then excludes models for which the number of photons expected from dark matter annihilation/decay
within the dwarf galaxy could not be accommodated by a $2\sigma$ downward statistical fluctuation in the number of foreground
photons within the set of energy bins which encompass the entire dwarf galaxy dark matter signal.  Since the sensitivity is controlled
by statistical uncertainties, it is proportional to $\sqrt{A_\textrm{eff} T_\textrm{obs} / \epsilon}$ for a line search,
where $A_\textrm{eff}$ is the effective area and $T_\textrm{obs}$ is the run time.

We thus see that a larger $A_\textrm{eff}$ and larger $T_\textrm{obs}$ would improve the
sensitivity of a dwarf galaxy search, while an improved energy resolution would improve the sensitivity of both
dwarf galaxy and diffuse emission searches.  The energy range, energy and angular resolutions ($1\sigma$), and effective area
of the relevant instruments are listed in Table~\ref{tab:experiments}.
Note that the listed values for the resolutions and effective areas are given at a particular energy, and our analysis is performed with these benchmark numbers.
If possible, we use values corresponding to photon energies at or near $100\mev$ with photon detection at normal incidence.
For the experiments that cover a lower energy range, we use the performance goal values.
In the case that only an estimated range is provided, we use the more pessimistic (worse resolution, smaller effective area) end of that range.
We list two benchmark points for ASTROGAM: one for photon detection via Compton scattering (below $10\mev$) and one for detection via pair production (above $10\mev$).

\begin{table}[h]
  \centering
  \begin{tabular}{cccccc}
  \hline
    Detector & Source & Energy Range [MeV] & $\epsilon$
    & PSF & $A_\textrm{eff}$ [$\cm^2$] \\
    \hline
    ACT & \cite{Boggs:2006mh} 
    & 0.2 -- 10 & 1\% & $1^\circ$ & 1000 \\
    GRIPS & \cite{Greiner:2011ih} 
    & 0.2 -- 80 & 3\% & $1.5^\circ$ & 200 \\
    AdEPT & \cite{Hunter:2013wla} 
    & 5 -- 200 & 15\% & $0.5^\circ$ & 600 \\
    COMPTEL & \cite{Weidenspointner:1999thesis,Kappadath:1998thesis} 
    & 0.8 -- 30 & 2\% & $2^\circ$ & 50  \\
    EGRET & \cite{Thompson:1993lr} 
    & 30 -- $10^4$ & 12.5\% & $2.8^\circ$ & 1000 \\
    Fermi-LAT & \cite{Atwood:2009ez} 
    & 20 -- $3\cdot 10^5$ & 7.5\% & $2^\circ$ & 4000 \\
    GAMMA-400 & \cite{Galper:2014pua} 
    & 100 -- $3\cdot 10^6$ & 12\% & $2^\circ$ & 3000 \\
    ASTROGAM (below $10\mev$) & \cite{astrogam} 
    & 0.3 -- 10 & 1\% & $0.5^\circ$ & 119 \\
    ASTROGAM (above $10\mev$) & \cite{astrogam} 
    & 10 -- 3000 & 30\% & $0.5^\circ$ & 514
  \end{tabular}
  \caption{Experimental parameters used to determine indirect detection bounds for dark matter annihilation and decay.
    The values between experiments are not directly comparable, since they may have been taken at different photon energies.
    The width of the point spread function (PSF) and $A_\textrm{eff}$
    values are not needed in the analysis for COMPTEL, EGRET, and Fermi, but they are included for completeness.}
  \label{tab:experiments}
\end{table}

\section{CURRENT BOUNDS AND FUTURE SENSITIVITY}

\begin{figure}[ht]
  \centerline{\includegraphics[width=0.95\textwidth]{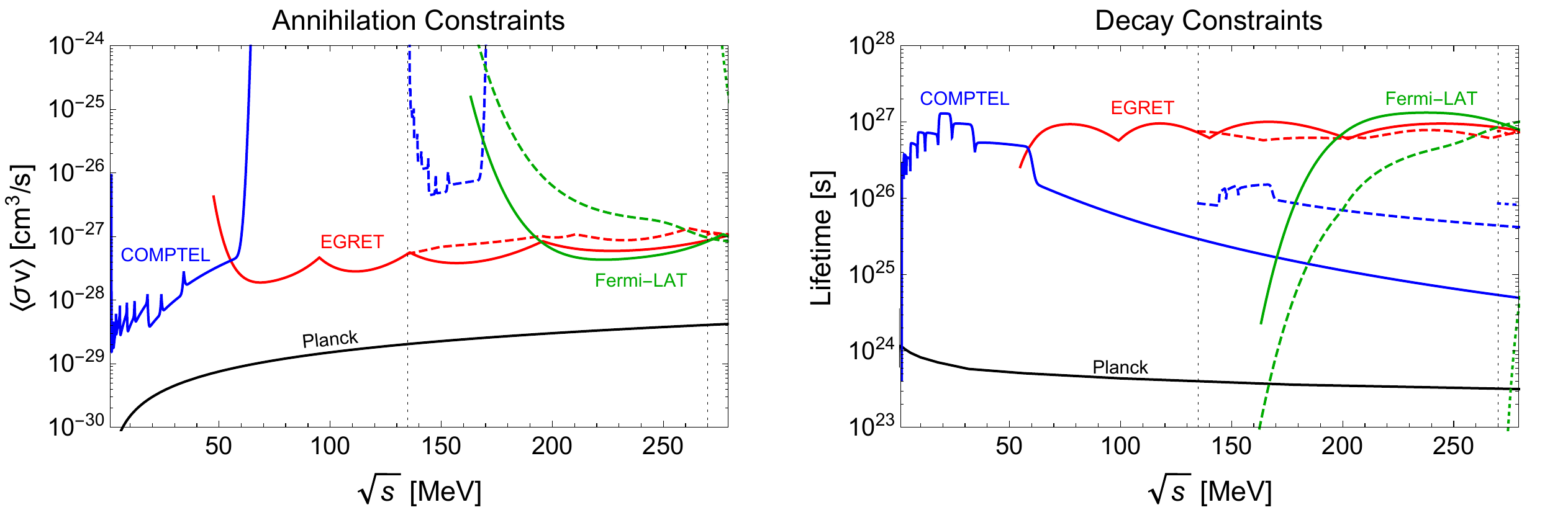}}
  \caption{Diffuse gamma ray flux constraints on dark matter annihilation (left panel) and decay (right panel) to the $\gamma \gamma$ (solid),
  $\gamma \pi^0$ (dashed), and $\pi^0 \pi^0$ (dotted) channels obtained from data from COMPTEL (blue), EGRET (red) and
  Fermi-LAT (green).  Also shown are constraints which can be obtained from Planck data (black).  The vertical
  black dashed lines are the kinematic thresholds for the $\gamma \pi^0$ and $\pi^0 \pi^0$ channels.}
\label{fig:constraints}
\end{figure}

In Figure~\ref{fig:constraints}, we plot current constraints on diffuse emission of photons from dark matter annihilation (left panel)
and decay (right panel) using data from COMPTEL, EGRET and Fermi-LAT, applying a conservative analysis (the details are described in~\cite{Boddy:2015efa}).
The $\gamma \gamma$, $\gamma \pi^0$ and $\pi^0 \pi^0$ channels are depicted with solid, dashed and dotted contours,
respectively.  Note that, although Fermi's energy range goes down as far as $20\mev$, their diffuse flux analysis only
extends down to $100\mev$, in order to reduce backgrounds from cosmic ray interactions with the atmosphere.
The $J$-factors used in estimating
the dark matter signal flux are obtained from~\cite{Cirelli:2010xx}.
We also plot bounds on the annihilation cross section (assuming $s$-wave annihilation) and decay rate~\cite{Slatyer} which arise
from measurements by Planck~\cite{Ade:2015xua} of the Cosmic Microwave Background (CMB) spectrum.  Note that,
although CMB constraints exceed those of indirect detection for the case of dark matter annihilation, the CMB
constraints are subleading for the case of dark matter decay.  This is not surprising, given that the
dark matter annihilation scales quadratically with dark matter density; the rate of dark matter annihilation
in the early Universe is thus greatly enhanced relative to the current epoch.

\begin{figure}[ht]
\begin{tabular}{c}
  \includegraphics[width=0.95\textwidth]{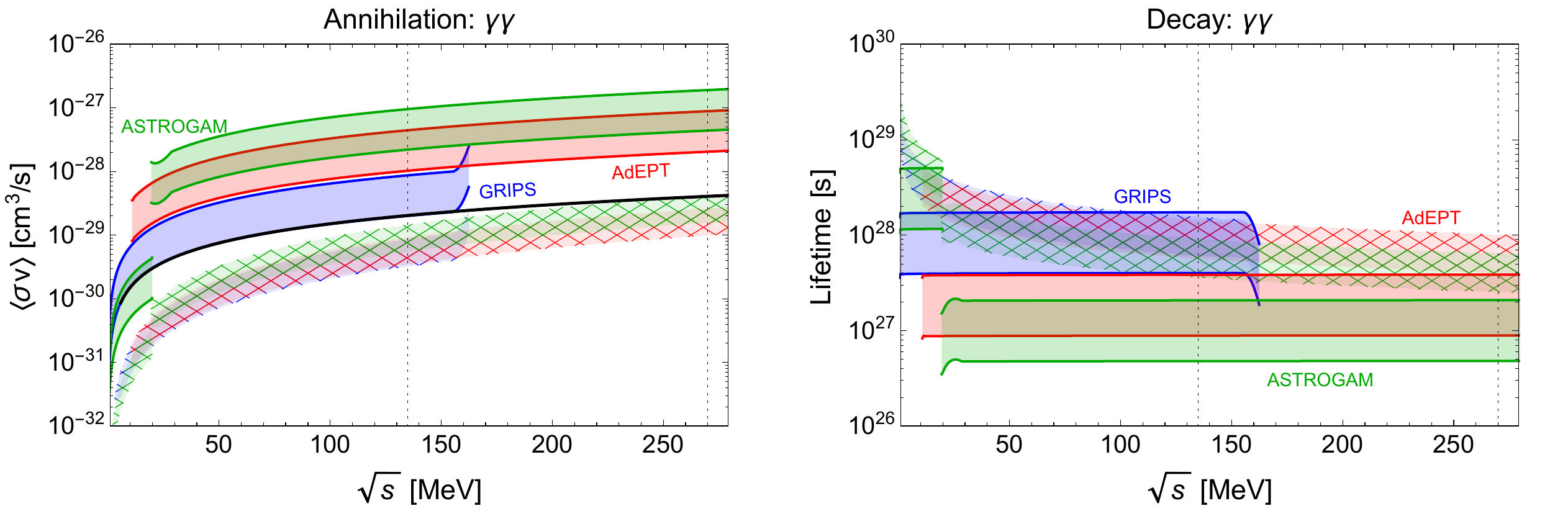} \\
  \includegraphics[width=0.95\textwidth]{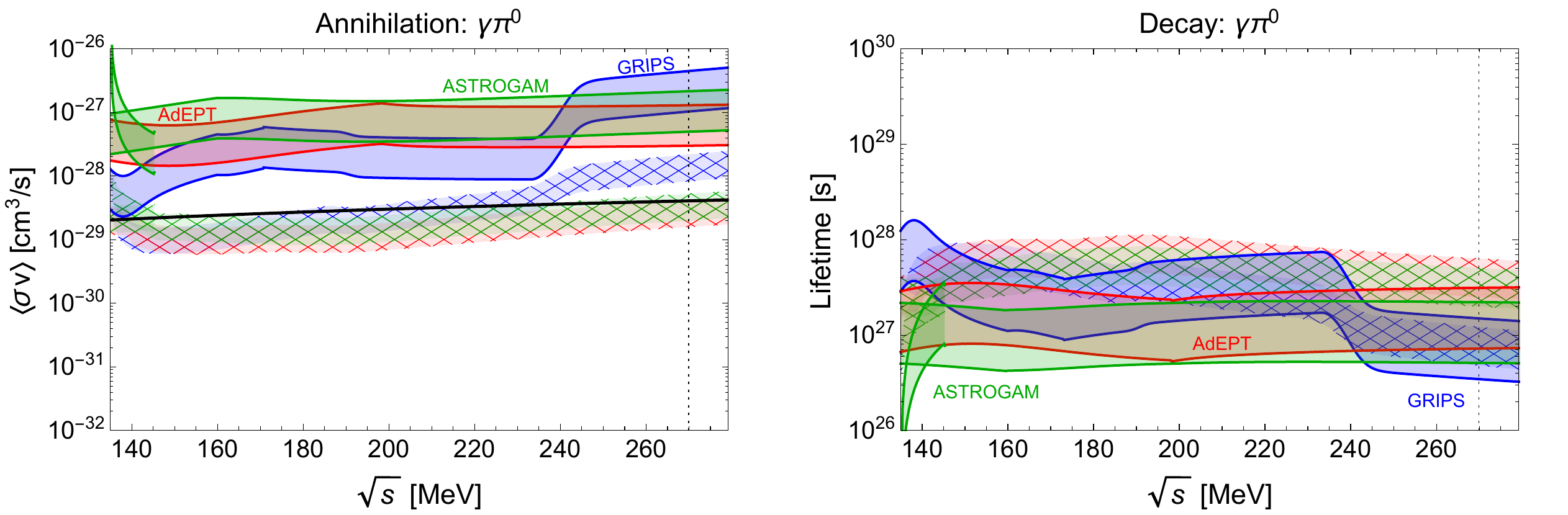} \\
  \includegraphics[width=0.95\textwidth]{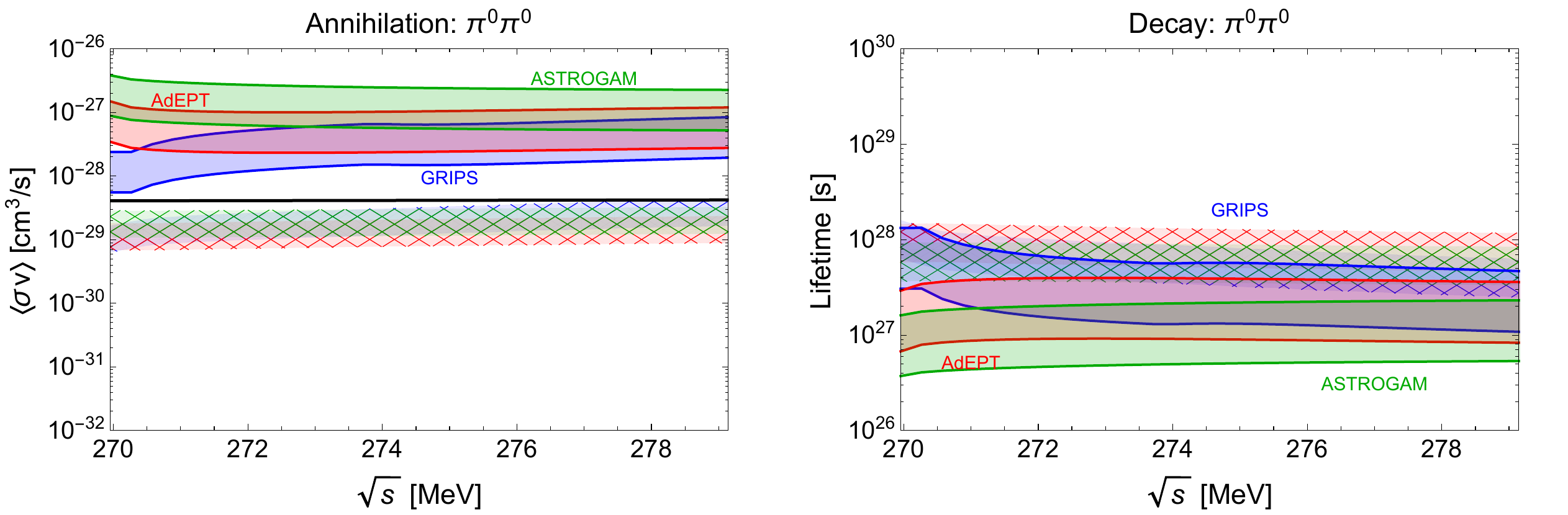} \\
\end{tabular}
    \caption{The projected sensitivity of ASTROGAM, AdEPT, and GRIPS to dark matter annihilation (left panels) or
    decay (right panels) in the $\gamma \gamma$ (top row), $\gamma \pi^0$ (middle row) and $\pi^0 \pi^0$ (bottom row)
    channels.  The shaded regions are bounded by contours which describe sensitivity of a conservative or
    optimistic analysis of diffuse gamma ray emission, as described in the text.  The hatched region describes the
    sensitivity contour for a search of Draco, when the $J$-factor is varied through $1\sigma$ systematic uncertainties.
    The black line describes the exclusion contour from Planck data.  The dashed vertical lines are the kinematic thresholds
    for the $\gamma \pi^0$ and $\pi^0 \pi^0$ channels.
    }
\label{fig:constraints-future}
\end{figure}

In Figure~\ref{fig:constraints-future}, we plot the sensitivity of a variety of future instruments, in the case
of either a diffuse emission search, or a 5-year search of photons from Draco (the $J$-factor from Draco is obtained
from~\cite{Geringer-Sameth:2014yza}).
A detailed description of the analysis is found in~\cite{Boddy:2015efa}.
The results of conservative and optimistic analyses of diffuse emission are
depicted as the boundaries of the solid bands, while the sensitivity of the analysis of Draco is depicted by
the hatched regions whose thicknesses represent the $1\sigma$ systematic uncertainties in $J$.

One can see several features in the sensitivities shown in Figure~\ref{fig:constraints-future}.
In the $\gamma\gamma$ and $\gamma\pi^0$ channels, ASTROGAM's Compton detector can detect the lower energy photons with a high energy resolution; consequently, it has the greatest sensitivity in the $\gamma\gamma$ channel in a diffuse gamma ray search.
Similarly, the enhanced sensitivity of GRIPS over AdEPT (and ASTROGAM in the pair-production regime) to the $\gamma \gamma$ channel in a diffuse gamma ray search is due to its better energy resolution.
ASTROGAM is the least sensitive for a diffuse search in its pair production regime, given the pessimistic projected 30\% energy resolution;
using a more optimistic 20\% would bring its sensitivities more in line with those of AdEPT.
However, AdEPT, GRIPS, and ASTROGAM all have similar sensitivity to the same channel in a search of dwarf galaxies because the greater effective area of AdEPT and ASTROGAM compensates for poorer energy resolution.
For the $\gamma \gamma$ and $\gamma \pi^0$
channels, the sensitivity of GRIPS is suppressed at higher energies due to the the relatively low upper limit of its energy range ($80\mev$),
compared to AdEPT and ASTROGAM.
Additionally, GRIPS will miss a fraction of the box-like feature arising from $\pi^0$ decay, unless the pion is nearly at rest.

Comparing the cases of dark matter annihilation and dark matter decay, one sees that a search of dwarf spheroidal galaxies shows a
greater improvement in sensitivity, relative to diffuse flux searches, for the case of dark matter annihilation.  This is not surprising, since
the enhanced dark matter density within a dwarf galaxy will lead to a quadratic enhancement in the annihilation rate, but only a linear
enhancement in the decay rate.

\section{CONCLUSIONS}

We have considered a class of models in which low-mass ($\sqrt{s} < 2m_{\pi^\pm}$) dark matter couples dominantly to quarks.  For these
models, kinematic considerations favor the $\gamma \gamma$, $\gamma \pi^0$ and $\pi^0 \pi^0$ final states for dark matter
annihilation or decay, regardless of the details of the interaction of dark matter with quarks.  These models thus generically
predict sharp and striking signals in the photons spectrum in the energy range ${\cal O}(1-100)\mev$.

This is especially interesting because, although there is currently a dearth of instruments focused on this
energy range, the astrophysics community is considering several plans to fill this ``MeV gap."  These new instruments
could improve sensitivity to these indirect detection channels by up to a few orders of magnitude.  Indeed, although
constraints from CMB observations are by far the most stringent current limits on dark matter annihilation, searches of
dwarf galaxies by future instruments could exceed those limits by more than an order of magnitude.
New instruments searching for photons in the ${\cal O}(1-100)\mev$ range may exceed previous instruments
in effective area, energy resolution and angular resolution, providing a variety of paths for improving sensitivity to low-mass
dark matter.

\section{ACKNOWLEDGMENTS}

This research is funded in part by NSF CAREER grant PHY-1250573.  JK thanks CETUP*
(Center for Theoretical Underground Physics and Related Areas), for its hospitality and partial support during the 2015 Summer Program.


\bibliographystyle{aipnum-cp}%
\bibliography{CETUP}%

\begin{thebibliography}{18}%
\makeatletter
\providecommand \@ifxundefined [1]{%
 \@ifx{#1\undefined}
}%
\providecommand \@ifnum [1]{%
 \ifnum #1\expandafter \@firstoftwo
 \else \expandafter \@secondoftwo
 \fi
}%
\providecommand \@ifx [1]{%
 \ifx #1\expandafter \@firstoftwo
 \else \expandafter \@secondoftwo
 \fi
}%
\providecommand \natexlab [1]{#1}%
\providecommand \enquote  [1]{``#1''}%
\providecommand \bibnamefont  [1]{#1}%
\providecommand \bibfnamefont [1]{#1}%
\providecommand \citenamefont [1]{#1}%
\providecommand \href@noop [0]{\@secondoftwo}%
\providecommand \href [0]{\begingroup \@sanitize@url \@href}%
\providecommand \@href[1]{\@@startlink{#1}\@@href}%
\providecommand \@@href[1]{\endgroup#1\@@endlink}%
\providecommand \@sanitize@url [0]{\catcode `\\12\catcode `\$12\catcode
  `\&12\catcode `\#12\catcode `\^12\catcode `\_12\catcode `\%12\relax}%
\providecommand \@@startlink[1]{}%
\providecommand \@@endlink[0]{}%
\providecommand \url  [0]{\begingroup\@sanitize@url \@url }%
\providecommand \@url [1]{\endgroup\@href {#1}{\urlprefix }}%
\providecommand \urlprefix  [0]{URL }%
\providecommand \Eprint [0]{\href }%
\providecommand \doibase [0]{http://dx.doi.org/}%
\providecommand \selectlanguage [0]{\@gobble}%
\providecommand \bibinfo  [0]{\@secondoftwo}%
\providecommand \bibfield  [0]{\@secondoftwo}%
\providecommand \translation [1]{[#1]}%
\providecommand \BibitemOpen [0]{}%
\providecommand \bibitemStop [0]{}%
\providecommand \bibitemNoStop [0]{.\EOS\space}%
\providecommand \EOS [0]{\spacefactor3000\relax}%
\providecommand \BibitemShut  [1]{\csname bibitem#1\endcsname}%
\let\auto@bib@innerbib\@empty
\bibitem [{\citenamefont {Boddy}\ and\ \citenamefont
  {Kumar}(2015)}]{Boddy:2015efa}%
  \BibitemOpen
  \bibfield  {author} {\bibinfo {author} {\bibfnamefont {K.~K.}\ \bibnamefont
  {Boddy}}\ and\ \bibinfo {author} {\bibfnamefont {J.}~\bibnamefont {Kumar}},\
  }\href@noop {} {\bibfield  {journal} {\bibinfo  {journal} {Phys. Rev.}\
  }\textbf {\bibinfo {volume} {D92}},\ p.\ \bibinfo {pages} {023533} (\bibinfo
  {year} {2015})},\ \Eprint {http://arxiv.org/abs/1504.04024} {arXiv:1504.04024
  [astro-ph.CO]} \BibitemShut {NoStop}%
\bibitem [{\citenamefont {Greiner}\ \emph {et~al.}(2012)\citenamefont {Greiner}
  \emph {et~al.}}]{Greiner:2011ih}%
  \BibitemOpen
  \bibfield  {author} {\bibinfo {author} {\bibfnamefont {J.}~\bibnamefont
  {Greiner}} \emph {et~al.},\ }\href@noop {} {\bibfield  {journal} {\bibinfo
  {journal} {Exper. Astron.}\ }\textbf {\bibinfo {volume} {34}},\ \unskip\
  \bibinfo {pages} {551--582} (\bibinfo {year} {2012})},\ \Eprint
  {http://arxiv.org/abs/1105.1265} {arXiv:1105.1265 [astro-ph.HE]} \BibitemShut
  {NoStop}%
\bibitem [{\citenamefont {Boggs}\ \emph {et~al.}(2006)\citenamefont {Boggs}
  \emph {et~al.}}]{Boggs:2006mh}%
  \BibitemOpen
  \bibfield  {author} {\bibinfo {author} {\bibfnamefont {S.~E.}\ \bibnamefont
  {Boggs}} \emph {et~al.} (\bibinfo {collaboration} {Larger ACT}),\ }\href@noop
  {} {\  (\bibinfo {year} {2006})},\ \Eprint
  {http://arxiv.org/abs/astro-ph/0608532} {arXiv:astro-ph/0608532 [astro-ph]}
  \BibitemShut {NoStop}%
\bibitem [{\citenamefont {{Buckley}}\ and\ \citenamefont {{APT
  Collaboration}}(2008)}]{Buckley:2008fk}%
  \BibitemOpen
  \bibfield  {author} {\bibinfo {author} {\bibfnamefont {J.~H.}\ \bibnamefont
  {{Buckley}}}\ and\ \bibinfo {author} {\bibnamefont {{APT Collaboration}}},\
  }\enquote {\bibinfo {title} {{The Advanced Pair Telescope (APT) Mission
  Concept}},}\ in\ \href@noop {} {\emph {\bibinfo {booktitle} {AAS/High Energy
  Astrophysics Division \#10}}},\ Vol.~\bibinfo {volume} {10}\ (\bibinfo {year}
  {2008})\ p.\ \bibinfo {pages} {\#37.04}\BibitemShut {NoStop}%
\bibitem [{\citenamefont {Hunter}\ \emph {et~al.}(2014)\citenamefont {Hunter}
  \emph {et~al.}}]{Hunter:2013wla}%
  \BibitemOpen
  \bibfield  {author} {\bibinfo {author} {\bibfnamefont {S.~D.}\ \bibnamefont
  {Hunter}} \emph {et~al.},\ }\href@noop {} {\bibfield  {journal} {\bibinfo
  {journal} {Astropart. Phys.}\ }\textbf {\bibinfo {volume} {59}},\ \unskip\
  \bibinfo {pages} {18--28} (\bibinfo {year} {2014})},\ \Eprint
  {http://arxiv.org/abs/1311.2059} {arXiv:1311.2059 [astro-ph.IM]} \BibitemShut
  {NoStop}%
\bibitem [{\citenamefont {Wu}\ \emph {et~al.}(2014)\citenamefont {Wu},
  \citenamefont {Su}, \citenamefont {Bravar}, \citenamefont {Chang},
  \citenamefont {Fan}, \citenamefont {Pohl},\ and\ \citenamefont
  {Walter}}]{Wu:2014tya}%
  \BibitemOpen
  \bibfield  {author} {\bibinfo {author} {\bibfnamefont {X.}~\bibnamefont
  {Wu}}, \bibinfo {author} {\bibfnamefont {M.}~\bibnamefont {Su}}, \bibinfo
  {author} {\bibfnamefont {A.}~\bibnamefont {Bravar}}, \bibinfo {author}
  {\bibfnamefont {J.}~\bibnamefont {Chang}}, \bibinfo {author} {\bibfnamefont
  {Y.}~\bibnamefont {Fan}}, \bibinfo {author} {\bibfnamefont {M.}~\bibnamefont
  {Pohl}}, \ and\ \bibinfo {author} {\bibfnamefont {R.}~\bibnamefont
  {Walter}},\ }\bibfield  {booktitle} {\emph {\bibinfo {booktitle}
  {{Proceedings, SPIE Astronomical Telescopes + Instrumentation 2014 :
  Ultraviolet to Gamma Ray}}},\ }\href@noop {} {\bibfield  {journal} {\bibinfo
  {journal} {Proc. SPIE Int. Soc. Opt. Eng.}\ }\textbf {\bibinfo {volume}
  {9144}},\ p.\ \bibinfo {pages} {91440F} (\bibinfo {year} {2014})},\ \Eprint
  {http://arxiv.org/abs/1407.0710} {arXiv:1407.0710 [astro-ph.IM]} \BibitemShut
  {NoStop}%
\bibitem [{\citenamefont {{Kierans}}\ \emph {et~al.}(2014)\citenamefont
  {{Kierans}}, \citenamefont {{Boggs}}, \citenamefont {{Lowell}}, \citenamefont
  {{Tomsick}}, \citenamefont {{Zoglauer}}, \citenamefont {{Amman}},
  \citenamefont {{Chiu}}, \citenamefont {{Chang}}, \citenamefont {{Lin}},
  \citenamefont {{Jean}}, \citenamefont {{von Ballmoos}}, \citenamefont
  {{Yang}}, \citenamefont {{Shang}}, \citenamefont {{Tseng}}, \citenamefont
  {{Chou}},\ and\ \citenamefont {{Chang}}}]{Kierans:2014lr}%
  \BibitemOpen
  \bibfield  {author} {\bibinfo {author} {\bibfnamefont {C.~A.}\ \bibnamefont
  {{Kierans}}}, \bibinfo {author} {\bibfnamefont {S.~E.}\ \bibnamefont
  {{Boggs}}}, \bibinfo {author} {\bibfnamefont {A.}~\bibnamefont {{Lowell}}},
  \bibinfo {author} {\bibfnamefont {J.}~\bibnamefont {{Tomsick}}}, \bibinfo
  {author} {\bibfnamefont {A.}~\bibnamefont {{Zoglauer}}}, \bibinfo {author}
  {\bibfnamefont {M.}~\bibnamefont {{Amman}}}, \bibinfo {author} {\bibfnamefont
  {J.-L.}\ \bibnamefont {{Chiu}}}, \bibinfo {author} {\bibfnamefont {H.-K.}\
  \bibnamefont {{Chang}}}, \bibinfo {author} {\bibfnamefont {C.-H.}\
  \bibnamefont {{Lin}}}, \bibinfo {author} {\bibfnamefont {P.}~\bibnamefont
  {{Jean}}}, \bibinfo {author} {\bibfnamefont {P.}~\bibnamefont {{von
  Ballmoos}}}, \bibinfo {author} {\bibfnamefont {C.-Y.}\ \bibnamefont
  {{Yang}}}, \bibinfo {author} {\bibfnamefont {J.-R.}\ \bibnamefont {{Shang}}},
  \bibinfo {author} {\bibfnamefont {C.-H.}\ \bibnamefont {{Tseng}}}, \bibinfo
  {author} {\bibfnamefont {Y.}~\bibnamefont {{Chou}}}, \ and\ \bibinfo {author}
  {\bibfnamefont {Y.-H.}\ \bibnamefont {{Chang}}},\ }\enquote {\bibinfo {title}
  {{Calibration of the Compton Spectrometer and Imager in preparation for the
  2014 balloon campaign}},}\ in\ \href@noop {} {\emph {\bibinfo {booktitle}
  {Space Telescopes and Instrumentation 2014: Ultraviolet to Gamma Ray}}},\
  \bibinfo {series} {Society of Photo-Optical Instrumentation Engineers (SPIE)
  Conference Series}, Vol.\ \bibinfo {volume} {9144}\ (\bibinfo {address}
  {Montreal, Quebec, Canada},\ \bibinfo {year} {2014})\ p.\ \bibinfo {pages}
  {91443M}\BibitemShut {NoStop}%
\bibitem [{\citenamefont {{ASTROGAM Collaboration}}()}]{astrogam}%
  \BibitemOpen
  \bibfield  {author} {\bibinfo {author} {\bibnamefont {{ASTROGAM
  Collaboration}}},\ }\href@noop {} {}\bibinfo {note}
  {{http://astrogam.iaps.inaf.it/}}\BibitemShut {NoStop}%
\bibitem [{\citenamefont {Strong}, \citenamefont {Moskalenko},\ and\
  \citenamefont {Reimer}(2004)}]{Strong:2004de}%
  \BibitemOpen
  \bibfield  {author} {\bibinfo {author} {\bibfnamefont {A.~W.}\ \bibnamefont
  {Strong}}, \bibinfo {author} {\bibfnamefont {I.~V.}\ \bibnamefont
  {Moskalenko}}, \ and\ \bibinfo {author} {\bibfnamefont {O.}~\bibnamefont
  {Reimer}},\ }\href@noop {} {\bibfield  {journal} {\bibinfo  {journal}
  {Astrophys. J.}\ }\textbf {\bibinfo {volume} {613}},\ \unskip\ \bibinfo
  {pages} {962--976} (\bibinfo {year} {2004})},\ \Eprint
  {http://arxiv.org/abs/astro-ph/0406254} {arXiv:astro-ph/0406254 [astro-ph]}
  \BibitemShut {NoStop}%
\bibitem [{\citenamefont {Weidenspointner}(1999)}]{Weidenspointner:1999thesis}%
  \BibitemOpen
  \bibfield  {author} {\bibinfo {author} {\bibfnamefont {G.}~\bibnamefont
  {Weidenspointner}},\ }\enquote {\bibinfo {title} {The origin of the cosmic
  gamma-ray background in the comptel energy range},}\ \href@noop {} {Ph.D.
  thesis},\ \bibinfo  {school} {Technical University of Munich}, \bibinfo
  {address} {Munich, Germany} \bibinfo {year} {1999}\BibitemShut {NoStop}%
\bibitem [{\citenamefont {Kappadath}(1998)}]{Kappadath:1998thesis}%
  \BibitemOpen
  \bibfield  {author} {\bibinfo {author} {\bibfnamefont {S.~C.}\ \bibnamefont
  {Kappadath}},\ }\enquote {\bibinfo {title} {Measurement of the cosmic diffuse
  gamma-ray spectrum from 800 kev to 30 mev},}\ \href@noop {} {Ph.D. thesis},\
  \bibinfo  {school} {University of New Hampshire}May \bibinfo {year}
  {1998}\BibitemShut {NoStop}%
\bibitem [{\citenamefont {{Thompson}}\ \emph {et~al.}(1993)\citenamefont
  {{Thompson}}, \citenamefont {{Bertsch}}, \citenamefont {{Fichtel}},
  \citenamefont {{Hartman}}, \citenamefont {{Hofstadter}}, \citenamefont
  {{Hughes}}, \citenamefont {{Hunter}}, \citenamefont {{Hughlock}},
  \citenamefont {{Kanbach}}, \citenamefont {{Kniffen}}, \citenamefont {{Lin}},
  \citenamefont {{Mattox}}, \citenamefont {{Mayer-Hasselwander}}, \citenamefont
  {{von Montigny}}, \citenamefont {{Nolan}}, \citenamefont {{Nel}},
  \citenamefont {{Pinkau}}, \citenamefont {{Rothermel}}, \citenamefont
  {{Schneid}}, \citenamefont {{Sommer}}, \citenamefont {{Sreekumar}},
  \citenamefont {{Tieger}},\ and\ \citenamefont {{Walker}}}]{Thompson:1993lr}%
  \BibitemOpen
  \bibfield  {author} {\bibinfo {author} {\bibfnamefont {D.~J.}\ \bibnamefont
  {{Thompson}}}, \bibinfo {author} {\bibfnamefont {D.~L.}\ \bibnamefont
  {{Bertsch}}}, \bibinfo {author} {\bibfnamefont {C.~E.}\ \bibnamefont
  {{Fichtel}}}, \bibinfo {author} {\bibfnamefont {R.~C.}\ \bibnamefont
  {{Hartman}}}, \bibinfo {author} {\bibfnamefont {R.}~\bibnamefont
  {{Hofstadter}}}, \bibinfo {author} {\bibfnamefont {E.~B.}\ \bibnamefont
  {{Hughes}}}, \bibinfo {author} {\bibfnamefont {S.~D.}\ \bibnamefont
  {{Hunter}}}, \bibinfo {author} {\bibfnamefont {B.~W.}\ \bibnamefont
  {{Hughlock}}}, \bibinfo {author} {\bibfnamefont {G.}~\bibnamefont
  {{Kanbach}}}, \bibinfo {author} {\bibfnamefont {D.~A.}\ \bibnamefont
  {{Kniffen}}}, \bibinfo {author} {\bibfnamefont {Y.~C.}\ \bibnamefont
  {{Lin}}}, \bibinfo {author} {\bibfnamefont {J.~R.}\ \bibnamefont {{Mattox}}},
  \bibinfo {author} {\bibfnamefont {H.~A.}\ \bibnamefont
  {{Mayer-Hasselwander}}}, \bibinfo {author} {\bibfnamefont {C.}~\bibnamefont
  {{von Montigny}}}, \bibinfo {author} {\bibfnamefont {P.~L.}\ \bibnamefont
  {{Nolan}}}, \bibinfo {author} {\bibfnamefont {H.~I.}\ \bibnamefont {{Nel}}},
  \bibinfo {author} {\bibfnamefont {K.}~\bibnamefont {{Pinkau}}}, \bibinfo
  {author} {\bibfnamefont {H.}~\bibnamefont {{Rothermel}}}, \bibinfo {author}
  {\bibfnamefont {E.~J.}\ \bibnamefont {{Schneid}}}, \bibinfo {author}
  {\bibfnamefont {M.}~\bibnamefont {{Sommer}}}, \bibinfo {author}
  {\bibfnamefont {P.}~\bibnamefont {{Sreekumar}}}, \bibinfo {author}
  {\bibfnamefont {D.}~\bibnamefont {{Tieger}}}, \ and\ \bibinfo {author}
  {\bibfnamefont {A.~H.}\ \bibnamefont {{Walker}}} (\bibinfo {collaboration}
  {EGRET}),\ }\href@noop {} {\bibfield  {journal} {\bibinfo  {journal}
  {Astrophys. J. Suppl.}\ }\textbf {\bibinfo {volume} {86}},\ \unskip\ \bibinfo
  {pages} {629--656}June (\bibinfo {year} {1993})}\BibitemShut {NoStop}%
\bibitem [{\citenamefont {{Atwood}}\ \emph {et~al.}(2009)\citenamefont
  {{Atwood}}, \citenamefont {{Abdo}}, \citenamefont {{Ackermann}},
  \citenamefont {{Althouse}}, \citenamefont {{Anderson}}, \citenamefont
  {{Axelsson}}, \citenamefont {{Baldini}}, \citenamefont {{Ballet}},
  \citenamefont {{Band}}, \citenamefont {{Barbiellini}},\ and\ \citenamefont
  {et~al.}}]{Atwood:2009ez}%
  \BibitemOpen
  \bibfield  {author} {\bibinfo {author} {\bibfnamefont {W.~B.}\ \bibnamefont
  {{Atwood}}}, \bibinfo {author} {\bibfnamefont {A.~A.}\ \bibnamefont
  {{Abdo}}}, \bibinfo {author} {\bibfnamefont {M.}~\bibnamefont {{Ackermann}}},
  \bibinfo {author} {\bibfnamefont {W.}~\bibnamefont {{Althouse}}}, \bibinfo
  {author} {\bibfnamefont {B.}~\bibnamefont {{Anderson}}}, \bibinfo {author}
  {\bibfnamefont {M.}~\bibnamefont {{Axelsson}}}, \bibinfo {author}
  {\bibfnamefont {L.}~\bibnamefont {{Baldini}}}, \bibinfo {author}
  {\bibfnamefont {J.}~\bibnamefont {{Ballet}}}, \bibinfo {author}
  {\bibfnamefont {D.~L.}\ \bibnamefont {{Band}}}, \bibinfo {author}
  {\bibfnamefont {G.}~\bibnamefont {{Barbiellini}}}, \ and\ \bibinfo {author}
  {\bibnamefont {et~al.}} (\bibinfo {collaboration} {Fermi-LAT}),\ }\href@noop
  {} {\bibfield  {journal} {\bibinfo  {journal} {Astrophys. J.}\ }\textbf
  {\bibinfo {volume} {697}},\ \unskip\ \bibinfo {pages} {1071--1102}June
  (\bibinfo {year} {2009})},\ \Eprint {http://arxiv.org/abs/0902.1089}
  {0902.1089 [astro-ph.IM]} \BibitemShut {NoStop}%
\bibitem [{\citenamefont {Galper}\ \emph {et~al.}(2014)\citenamefont {Galper}
  \emph {et~al.}}]{Galper:2014pua}%
  \BibitemOpen
  \bibfield  {author} {\bibinfo {author} {\bibfnamefont {A.~M.}\ \bibnamefont
  {Galper}} \emph {et~al.},\ }\href@noop {} {\  (\bibinfo {year} {2014})},\
  \Eprint {http://arxiv.org/abs/1412.4239} {arXiv:1412.4239 [physics.ins-det]}
  \BibitemShut {NoStop}%
\bibitem [{\citenamefont {Cirelli}\ \emph {et~al.}(2011)\citenamefont
  {Cirelli}, \citenamefont {Corcella}, \citenamefont {Hektor}, \citenamefont
  {Hutsi}, \citenamefont {Kadastik}, \citenamefont {Panci}, \citenamefont
  {Raidal}, \citenamefont {Sala},\ and\ \citenamefont
  {Strumia}}]{Cirelli:2010xx}%
  \BibitemOpen
  \bibfield  {author} {\bibinfo {author} {\bibfnamefont {M.}~\bibnamefont
  {Cirelli}}, \bibinfo {author} {\bibfnamefont {G.}~\bibnamefont {Corcella}},
  \bibinfo {author} {\bibfnamefont {A.}~\bibnamefont {Hektor}}, \bibinfo
  {author} {\bibfnamefont {G.}~\bibnamefont {Hutsi}}, \bibinfo {author}
  {\bibfnamefont {M.}~\bibnamefont {Kadastik}}, \bibinfo {author}
  {\bibfnamefont {P.}~\bibnamefont {Panci}}, \bibinfo {author} {\bibfnamefont
  {M.}~\bibnamefont {Raidal}}, \bibinfo {author} {\bibfnamefont
  {F.}~\bibnamefont {Sala}}, \ and\ \bibinfo {author} {\bibfnamefont
  {A.}~\bibnamefont {Strumia}},\ }\href@noop {} {\bibfield  {journal} {\bibinfo
   {journal} {JCAP}\ }\textbf {\bibinfo {volume} {1103}},\ p.\ \bibinfo {pages}
  {051} (\bibinfo {year} {2011})},\ \bibinfo {note} {[Erratum:
  JCAP1210,E01(2012)]},\ \Eprint {http://arxiv.org/abs/1012.4515}
  {arXiv:1012.4515 [hep-ph]} \BibitemShut {NoStop}%
\bibitem [{\citenamefont {Slatyer}(2013)}]{Slatyer}%
  \BibitemOpen
  \bibfield  {author} {\bibinfo {author} {\bibfnamefont {T.~R.}\ \bibnamefont
  {Slatyer}},\ }\href@noop {} {\bibfield  {journal} {\bibinfo  {journal} {Phys.
  Rev.}\ }\textbf {\bibinfo {volume} {D87}},\ p.\ \bibinfo {pages} {125513}
  (\bibinfo {year} {2013})},\ \Eprint {http://arxiv.org/abs/1211.0283}
  {arXiv:1211.0283 [astro-ph.CO]} \BibitemShut {NoStop}%
\bibitem [{\citenamefont {Ade}\ \emph {et~al.}(2015)\citenamefont {Ade} \emph
  {et~al.}}]{Ade:2015xua}%
  \BibitemOpen
  \bibfield  {author} {\bibinfo {author} {\bibfnamefont {P.~A.~R.}\
  \bibnamefont {Ade}} \emph {et~al.} (\bibinfo {collaboration} {Planck}),\
  }\href@noop {} {\  (\bibinfo {year} {2015})},\ \Eprint
  {http://arxiv.org/abs/1502.01589} {arXiv:1502.01589 [astro-ph.CO]}
  \BibitemShut {NoStop}%
\bibitem [{\citenamefont {Geringer-Sameth}, \citenamefont {Koushiappas},\ and\
  \citenamefont {Walker}(2015)}]{Geringer-Sameth:2014yza}%
  \BibitemOpen
  \bibfield  {author} {\bibinfo {author} {\bibfnamefont {A.}~\bibnamefont
  {Geringer-Sameth}}, \bibinfo {author} {\bibfnamefont {S.~M.}\ \bibnamefont
  {Koushiappas}}, \ and\ \bibinfo {author} {\bibfnamefont {M.}~\bibnamefont
  {Walker}},\ }\href@noop {} {\bibfield  {journal} {\bibinfo  {journal}
  {Astrophys. J.}\ }\textbf {\bibinfo {volume} {801}},\ p.~\bibinfo {pages}
  {74} (\bibinfo {year} {2015})},\ \Eprint {http://arxiv.org/abs/1408.0002}
  {arXiv:1408.0002 [astro-ph.CO]} \BibitemShut {NoStop}%
\end{thebibliography}%

\end{document}